\documentclass[aps,prl,groupedaddress]{revtex4}
\usepackage{amsmath}
\usepackage{epsfig,epsf}
\usepackage{graphicx}
\usepackage{verbatim}
\usepackage{epstopdf}

\def\beq{\begin{equation}}
\def\eeq{\end{equation}}
\def\bea{\begin{eqnarray}}
\def\eea{\end{eqnarray}}
\def\eq#1{{Eq.~(\ref{#1})}}
\def\fig#1{{Fig.~\ref{#1}}}

\newcommand{\bas}{\bar{\alpha}_s}
\newcommand{\as}{\alpha_s}
\newcommand{\Lb}{\left(}
\newcommand{\Rb}{\right)}
\newcommand{\nn}{\nonumber}

\begin{document}


\voffset1.5cm
\title{Gluon saturation and inclusive production at low transverse momenta }
\author{Eugene Levin}
\affiliation{
Departamento de F\'\i sica, Universidad T\'ecnica
Federico Santa Mar\'\i a, Avda. Espa\~na 1680,
Casilla 110-V,  Valparaiso, Chile \\
 Department of Particle Physics,  Tel Aviv University , Tel Aviv 69978, Israel}
 
\date{\today}
\begin{abstract}
In this letter we suggest the generalization of $k_T$-factorization formula for inclusive gluon production for the dense-dense parton system scattering. It turnes out  that the soft gluon production with transverse momentum $p_T$  is suppressed by additional 
Sudakov-like factor that depends on $p^2_T/Q^2_s$ ratio in a good agreement with the first numerical calculation in Colour Glass Condensate approach by J.~P.~Blaizot, T.~Lappi and Y.~Mehtar-Tan.
\end{abstract}
\maketitle


It is well known that  our approach to inclusive production of  an gluon jet is based on $k_T$ factorization\cite{KTF1,KTF2,KTF3,KTF4}
which leads to

\beq \label{F1}
\frac{d \sigma}{d y \,d^2 p_{T}}\,\,=\,\,\frac{2 \pi \bas}{p^2_\perp}\int d^2 k_T\,\,\phi^{h_1}_G\Lb x_1;\vec{k}_T\Rb\,\phi^{h_2}_G\Lb x_2;\vec{p}_T -\vec{k}_T\Rb
\eeq
where $\phi^{h_i}_G$ are the probability to find a gluon that carries $x_i$ fraction of energy with $k_\perp$ transverse momentum and $ \bas \,= \,\as N_c/\pi$ with the number of colours equals $N_c$.

In the framework of high density QCD\cite{GLR,MUQI,MV,B,MUCD,K,JIMWLK,KLN,KLNLHC} the $k_T$ -factorization  has been proven \cite{KTINC} (see also Refs. \cite{BRINC,CMINC,KLINC,LPINC,KLPINC})
 for the scattering of the diluted system of partons, say for virtual photon, with the dense one.
Such scattering is characterized by two scale of hardness: the saturation momentum of the dense system $Q_s$ and the $p_T$ of the produced gluon. The dense-dense parton system scattering has three scales of hardness: two saturation momenta and $p_T$; and the $k_T$ -factorization has not been proven for this process. The most dangerous region is for $p_T$ smaller than both saturation momenta ( $p_t \,<\, Q_{1,s} \,\leq\,Q_{2s}$) where we did not expect that $k_T$ factorization will work. However, for $ Q_{1,s} \,<\,p_T \,<\,Q_{2s}$ we are dealing with scattering with two scales of hardness and we can expect that $k_T$ factorization is valid here.   In this paper we address this problem and suggests the generalization of the $k_T$ factorization (see below \eq{GGT}) for  $p_t \,<\, Q_{1,s} \,\leq\,Q_{2s}$.

As it was shown in Ref.\cite{KTINC}  $\phi^{h_i}_G\Lb x_1;\vec{k}_\perp\Rb$  can be written through  dipole scattering amplitude $N\Lb x_i, r_\perp; b \Rb$, where $  r_\perp$ is the dipole size and $b$ is the impact parameter of the scattering. This relation reads as follows
\beq \label{F2}
\phi^{h_i}_G\Lb x_i;\vec{k}_T\Rb\,\,=\,\,\frac{1}{\bas\,4 \pi}\,\int d^2 b \,d^2 r_\perp
e^{i \vec{k}_T\cdot \vec{r}_\perp}\,\,\nabla^2_\perp\,N^{h_i}_G\Lb y_i = \ln(1/x_i); r_\perp; b \Rb
\eeq
where 
\beq \label{F3}
N^{h_i}_G\Lb y_i = \ln(1/x_i); r_\perp; b \Rb\,\,=\,\,2 \,N\Lb y_i = \ln(1/x_i); r_\perp; b \Rb\,\,\,-\,\,\,N^2\Lb y_i = \ln(1/x_i); r_\perp; b \Rb
\eeq
$N\Lb y_i = \ln(1/x_i); r_\perp; b \Rb$ is the dipole -hadron ($h_i$) scattering amplitude which satisfy the Balitsky-Kovchegov equation.

Using that $N\Lb x_i, r_\perp; b \Rb$ is a function of $r^2_\perp$ we can rewrite \eq{F2} in the form
\bea 
\phi^{h_i}_G\Lb x_i;\vec{k}_T\Rb\,\,&=&\,\,\frac{1}{\bas\,4 \pi}\,\int d^2 b \,d^2 r_\perp
e^{i \vec{k}_T\cdot \vec{r}_\perp} \frac{1}{r_\perp}\,\frac{\partial}{\partial r_\perp }\,r_\perp \frac{\partial}{\partial r_\perp}\,N^{h_i}_G\Lb y_i = \ln(1/x_i); r_\perp; b \Rb;
 \,\label{41}\\
&=& \,\,\frac{1}{\bas\,4 \pi}\,\int d^2 b \,\Big\{ 2 \pi \Big(e^{i \vec{k}_T\cdot \vec{r}_\perp}\,\,\,r_\perp \frac{\partial}{\partial r_\perp}\,N^{h_i}_G\Lb y_i = \ln(1/x_i); r_\perp; b \Rb\Big)|^{r_\perp =\infty}_{r_\perp = 0}\,\,\nonumber\\
&-&\,\,i k_T\int d r_\perp \,d \phi \,\cos\phi e^{i k_T r_\perp\,\cos \phi}
\,r_\perp \frac{\partial}{\partial r_\perp}\,N^{h_i}_G\Lb y_i = \ln(1/x_i); r_\perp; b \Rb\Big\};\label{42}
\eea
At $r_\perp \to 0$ the amplitude $N_G $ approaches the solution of the DGLAP equation which in our case corresponds the double log limit of the BFKL equation. Therefore  
\bea \label{F5}
&&\int d^2 b\, r_\perp \frac{\partial}{\partial r_\perp}\,N^{h_i}_G\Lb y_i = \ln(1/x_i); r_\perp; b \Rb \,\to\,\mbox{DLA} \,\\
&&\,\propto \,r_\perp \frac{\partial}{\partial r_\perp}\,\exp\Big( \sqrt{4 \bas \ln(1/x)\,\ln(1/(r^2_\perp \lambda^2_{QCD})}\,-\, \,\,
\ln(1/(r^2_\perp \lambda^2_{QCD})\Big)\,\,\xrightarrow{r_\perp \to 0}\,0\nn
\eea
It is worth to mention that the impact parameter dependence  enters in \eq{F5} as a factor and cannot change our claim that this term vanishes at $r_\perp \to 0$.

At large $r_\perp$ $ 1 - N_G \propto \exp\Big( - C \ln^2\Lb r^2 Q^2_s \Rb\Big)$ (see Ref.\cite{LT}) and, therefore,

\beq \label{F6}
\int d^2 b\, r_\perp \frac{\partial}{\partial r_\perp}\,N^{h_i}_G\Lb y_i = \ln(1/x_i); r_\perp; b \Rb \,\xrightarrow{r_\perp \to \infty}\,\,0
\eeq

Finally  we see that the first term in \eq{42} vanishes and we have
\beq \label{F7}
\phi^{h_i}_G\Lb x_i;\vec{k}_T\Rb\,\,=\,\,\, k_T\int d r_\perp \,J_1\Lb k_T r_\perp\Rb\,
\,r_\perp \frac{\partial}{\partial r_\perp}\,N^{h_i}_G\Lb y_i = \ln(1/x_i); r_\perp; b \Rb
 \eeq

\begin{figure}[t]
\vspace{0.5cm}
\begin{minipage}{7cm}{
\includegraphics[width=6cm]{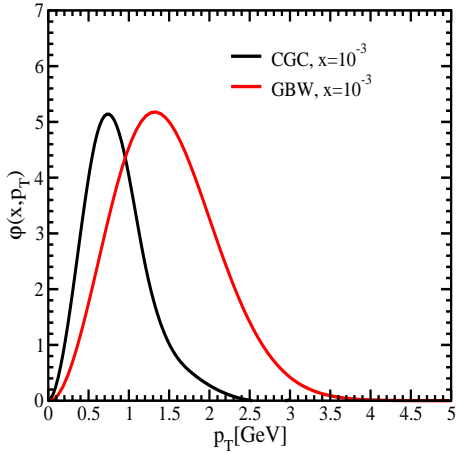}
}
\end{minipage}
\begin{minipage}{8cm}{
\caption{Function $\phi_G$ versus transverse momentum in different saturation models: GBW denotes the Golec-Biernat and Wusthoff model (see Ref.\cite{GBW}) while CGC denotes the model suggested in Ref.\cite{CGC}. These two model have different values of the saturation momentum and the picture illustrates that $\phi(x,p_T)$  has maximum at $p_T= Q_s$.
}
\label{num}
}
\end{minipage}
\end{figure}

From \eq{F7} one can see that at $k _T\,\to\, 0$  $\phi_G \,\propto \, k^2_T$  while at large values of $k_T$ 
$\phi_G 
\, \propto \, 1/k^2_T \to 0 $.  Such behaviour of $\phi_G$ means that it has maximum at $k_T \approx Q_s$.
The numerical calculations\footnote{ We are very thankful to Amir Rezaeian  who made this plot}(see \fig{num}) confirm this claim.

~
~

Having this feature of $\phi_G$ in mind we see that at $p_T \ll Q_s$ \eq{F1} gives
\beq \label{F8}
\frac{d \sigma}{d y \,d^2 p_{T}}\,\,=\,\,\frac{2 \pi \bas}{p^2_\perp}\int d^2 k_T\,\,\phi^{h_1}_G\Lb x_1;\vec{k}_T\Rb\,\phi^{h_2}_G\Lb x_2;\
\vec{k}_T\Rb
\eeq

One can see that at $p_T \to 0$ the cross section tends to infinity. Since we are talking about inclusive cross section 
generally speaking such situation is possible and it corresponds to increasing  multiplicity of soft gluons. However, in 
framework of gluon saturation it looks strange. As we have discussed above, the main contribution to $\phi_G$ give the gluons with transverse 
momenta of about $Q_s$ while the gluons with small values of $k_T$ are suppressed. In other words, the correlation length between emitted gluon is of the order of $1/Q_s$  and we expect that emission of gluons with the wave length larger that $1/Q_s$ should be suppressed.  Of course, soft gluons with $p_T \ll Q_s$  could be  emitted in the final state but they will not propagate through the medium since the cross section is large ($\sigma T \Lb b \Rb \,\propto \,Q^2_s/p^2_T \gg 1$).

\begin{figure}[ht]
\includegraphics[width=15cm]{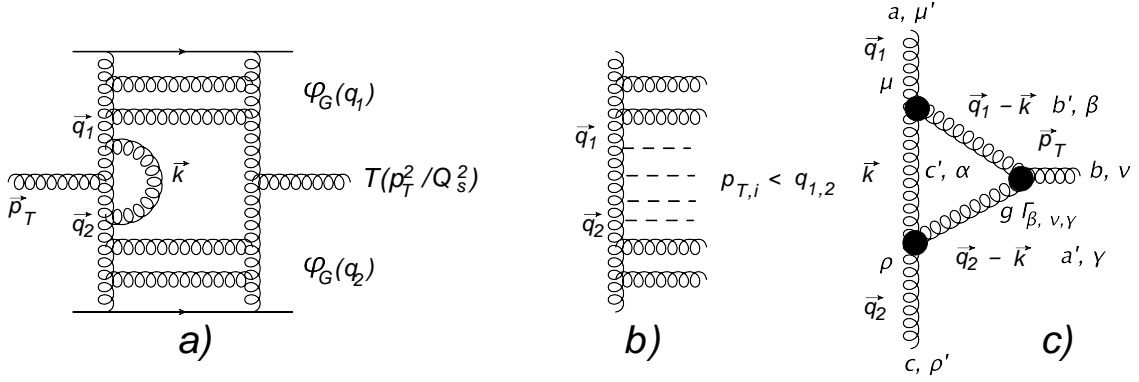}
\caption{Inclusive cross section: $\vec{q}_1 - \vec{q}_2 = \vec{p}_T$.
}
\label{GGT}
\end{figure}

In this letter we will show that simple formula of \eq{F1} should be changed and a new double log suppression factor ($T$) should be added. Therefore, the inclusive cross section has a form (see \fig{GGT}-a)
\beq \label{F9}
\frac{d \sigma}{d y \,d^2 p_{T}}\,\,=\,\,\frac{2 \pi \bas}{p^2_\perp}\int d^2 q_1\,d^2 q_2 \,\delta\Lb \vec{q}_1 - \vec{q}_2  - \vec{p}_T\Rb\,\,\phi^{h_1}_G\Lb x_1;\vec{q}_1\Rb\,\phi^{h_2}_G\Lb x_2;\vec{q}_2\Rb\,T\Big(\frac{p^2_T}{Q^2_s}\Big)
\eeq
In \eq{F9} we assume that $Q_{s,1} \approx Q_{s,2}$.
The appearance of $T$ in inclusive production  was found in 1980's \cite{DDT,PAPE,PASR} and it  is related to the fact that the emission of some gluons is suppressed in the process. In our case the emission of gluons is suppressed  with the value of the transverse momenta ( $p_{T,i}$) in the region: $ p_T\,\leq\,p_{T,i} \,\leq\,Q_s$ (see \fig{GGT}-b, where the gluons which emission is suppressed,  are denoted by the dashed lines).  Actually, the emission of gluons with small values of $p_{T,i}$ has been taken into account in functions $\phi^{h_i}_G$ but they result in suppression of the emission for  such gluons and we do not need to account separately for them.

\eq{F9} says that the emission of the gluon with $p_T \leq Q_s$ is suppressed and only gluons with $p_t > Q_s$ gives the contribution to the inclusive production. For such gluons the $k_T$ factorization works.
This qualitative features of \eq{F9} has been confirmed by the first numerical calculculation that found the deviation from $k_T$ factorizxation (see Ref.\cite{BLMT}). These calculation shows that for $p_T < Q_s$  the gluon production is suppressed while for $p_T > Q_s$ the  $k_T$ factorization works perfectly well.

We calculate the first diagrams of \fig{GGT}-c to illustrate the double log contribution.

This diagram is equal to
\bea \label{DI}
&&A\Lb \fig{GGT}-c\Rb\,\,=\,\,\frac{g^3}{ (2 \pi)^4\,i }\,\Big( f_{a,b',c'}\,f_{c',b,a}\,f_{a',c,c'}\,=\,-\,\frac{N_c}{2}\,f_{a b c}\Big)\,\int\,\frac{d k^+\,d k^- \,d^2 k_T}{2\,q^+_1 \cdot q^-_2} \\ &&\frac{q^+_{1,\mu}\Gamma_{\mu, \alpha,\beta}\, \Gamma_{\beta,\gamma,\nu}\, \Gamma_{\gamma,\alpha,\rho}\,q^-_{2,\rho}}{\Lb k^+k^-\,- k^2_T\,-\,i\epsilon\Rb\,\Lb (q^+_1 - k^+)(q^-_1 - k^-) \,-\,(\vec{q}_1 - \vec{k})^2_T\,-\,i\epsilon\Rb\,\Lb (q^+_2 - k^+)(q^-_2 - k^-) \,-\,(\vec{q}_2 - \vec{k})^2_T\,-\,i\epsilon\Rb} \nn
\eea

In \eq{DI} we used that at high energies the propagators of gluons with momenta $q_1$ and $q_2$ can be written in the form \cite{BFKL,GLR}
\beq \label{PROPO}
D_{\mu',\mu}\Lb q_1\Rb\,\,=\,\,\frac{q^-_{2, \mu'}\,q^+_{1, \mu}}{ q^-_{2}\,q^+_{1}}\,\frac{1}{q^2_{1,T}};\,\,\,\,\,
D_{\rho,\rho'}\Lb q_2\Rb\,\,=\,\,\frac{q^-_{2, \rho}\,q^+_{1, \rho'}}{ q^-_{2}\,q^+_{1}}\,\frac{1}{q^2_{2,T}};
\eeq
In leading log(1/x) approximation we have the following kinematic constraints:
\beq \label{KICO}
q^+_1 \,\gg\,q^+_2;\,\,\,\,q^-_2 \,\gg\,q^-_1;\,\,\,\,q^+_1\,q^-_2\,=\,p^2_T; \,\,q^+_1\,q^-_1\,\,\ll\,\,q^2_{1,T};\,\,\,
\,\,q^+_2\,q^-_2\,\,\ll\,\,q^2_{2,T};
\eeq

Integration for $k_T\, > \,Q_s$ leads to renormalization of the coupling QCD constant. Therefore, we are interested in the kinematic region where
\beq \label{KRKT}
q_{i,T} \,\approx\,Q_s\,\gg\, k_T \,\gg\,p_T
\eeq

 Having in mind \eq{KICO} and \eq{KRKT}  we can rewrite \eq{DI} using
\beq \label{DI1}
\Lb q_1 - k\Rb ^2\,\,+\,i\epsilon =\,\, (k^-  -  q^-_1)\,(k^+ - q^+_1 )\,\,-\,\, Q^2_s \,-\,i\epsilon;\,\,\,
\Lb q_2 - k\Rb ^2\,\,+\,i\epsilon =\,\, (k^+ - q^-_2)  (k^- - q^-_2 )\,\,-\,\, Q^2_s \,-\,i\epsilon;
\eeq

One can see that for $ q^-_1\,<\,k^-\,<\,q^-_2$ the poles in $k^+$  in \eq{DI} are situated in diffrent semiplanes and we can close the contour in complex plane $k^+$ on the pole  from $\Lb q_2 - k\Rb ^2 = 0$. It gives
\beq \label{C1}
k^+_0\,\,=\,\,\frac{Q^2_s}{k^-\,-\,q^-_2}\,\,\to \,\,-\,\frac{Q^2_s}{q^-_2}
\eeq

\beq \label{DI3}
A\Lb \fig{GGT}-c\Rb\,\,=\,\,-\,\frac{g^3 \pi}{ (2 \pi)^3 }\,\,\frac{N_c}{2}\,f_{a b c}\,\int\,\frac{d k^2_T\,d k^-}{2\,p^2_T\,q^-_2 }\, \frac{q^+_{1,\mu}\Gamma_{\mu, \alpha,\beta}\, \Gamma_{\beta,\gamma,\nu}\, \Gamma_{\gamma,\alpha,\rho}\,q^-_{2,\rho}}{\Lb q^+_1\,k^- \,+\,Q^2_s\Rb\,\Lb k^- \frac{Q^2_s}{q^-_2} \,+\,k^2_T\Rb} 
\eeq
One can see that from the kinematic region given by
\beq \label{DLAKR}
\frac{k^2_T\,q^-_2}{Q^2_s}\,\,\, \gg\,\,\,  k^-\,\,\,\gg\,\,\,Q^2_s/q^+_1 ;\,\,\,\,\,\,\,Q^2_s\,\,\gg\,\,k^2_T\,\,\gg\,\,p^2_T
\eeq
we have a double log contribution, namely,

\beq \label{DI4}
A\Lb \fig{GGT}-c\Rb\,\,=\,\,-\,\frac{g^3\pi}{ (2 \pi)^3 }\,\,\frac{N_c}{2}\,f_{a b c}\,\frac{1}{4\,p^2_T\,(q^+_1\, q^-_2)} \,q^+_{1,\mu}\Gamma_{\mu, \alpha,\beta}\, \Gamma_{\beta,\gamma,\nu}\, \Gamma_{\gamma,\alpha,\rho}\,q^-_{2,\rho}\,\,
\ln^2\Big(Q^2_s/p^2_T\Big)
\eeq
 Direct calculation  of sum over gluon polarization in \eq{DI4}  leads to
\beq \label{DI5}
\,\frac{1}{p^2_T}\,q^+_{1,\mu}\Gamma_{\mu, \alpha,\beta}\, \Gamma_{\beta,\gamma,\nu}\, \Gamma_{\gamma,\alpha,\rho}\,q^-_{2,\rho}\,\,=\,\,
4\,(q^+_1\, q^-_2)\,\Gamma_\nu
\eeq
where 
\beq \label{DI6}
 \Gamma_\nu\,\,=\,\, 2\,\Big(\vec{q}_{T,1}\,\,-\,\,\frac{\vec{p}_{T}}{q^2_{T,1}}\Big)_\nu
\eeq
is famous BFKL vertex \cite{BFKL}.

Collecting all factors and using the notation $ \Gamma^{a b c}_\nu\,\,=\,\,2 g\,f_{a b c }\,\Gamma_\nu$ we obtain for the diagram of \fig{GGT}-c the following expression
\beq \label{DI7}
A\Lb \fig{GGT}-c\Rb\,\,=\,\,-\,\frac{\bas}{ 4 }\,\Gamma^{a b c}_\nu\,\ln^2\Big(Q^2_s/p^2_T\Big)
\eeq
where $\bas\,\,=\,\,\as N_c/\pi$.
Using the well known technique (see Refs.\cite{DDT,PAPE}) we obtain
\beq \label{TFA}
T\Big(\frac{p^2_T}{Q^2_s}\Big)\,\,\,=\,\,\,\exp\Big\{- \frac{\bas}{ 2 }\,\ln^2\Big(Q^2_s/p^2_T\Big) \Big\}
\eeq

This result follows directly from the generalization of Low theorem for soft photon\cite{LOW} for high energy scattering (Gribov's theorem \cite{GRIB}). It says that if $p_T$ of emitted photon smaller than any typical transverse moneta in the process the cross section of emitted photon is equal to
\beq \label{DI8}
\sigma_\gamma\,\,\,=\,\,\frac{\alpha}{2 \pi} \,\frac{d \omega}{\omega} \,\frac{ d^2 p_T}{p^2_T}\,\sigma_0
\eeq
where $ \sigma_0$ is the cross section for the process without photon. This theorem has been generalized to emission of gluons (see Ref. \cite{LIPLT}).  In our case the typical momentum scales of the process are $Q_{s,1}$ and $Q_{s,2}$ and, therefore,  gluons with 
$p_T \,\ll\,Q_{s,1} \,\mbox{and} \,Q_{s,2}$  are emitted independently according to the Poisson distribution. The emission of one gluon
will be suppressed by $\exp\Lb - \langle n \rangle \Rb$ where $\langle n \rangle\,=\,\frac{\bas}{ 2 }\,\ln^2\Big(Q^2_s/p^2_T\Big) $ is the average number of emitted gluons. In the case of two different saturations scales $Q_{s,1}$ and $Q_{s,2}$ the average multiplicity  $\langle n \rangle\,=\,\frac{\bas}{ 2 }\,\ln^2\Big(Q^2_{s,min}/p^2_T\Big)
$ where $Q_{s, min} = \mbox{min}\{Q_{s,1}, Q_{s,2}\}$.  For dilute-dense scattering the value of $Q_{s,min} $ vanishes and we have no additional suppression. This feature is related to the fact that the BFKL evolution has two branches one of which leads to decrease of the typical transverse momenta of gluons.

\begin{figure}[t]
\begin{minipage}{9cm}{
\includegraphics[width=8cm]{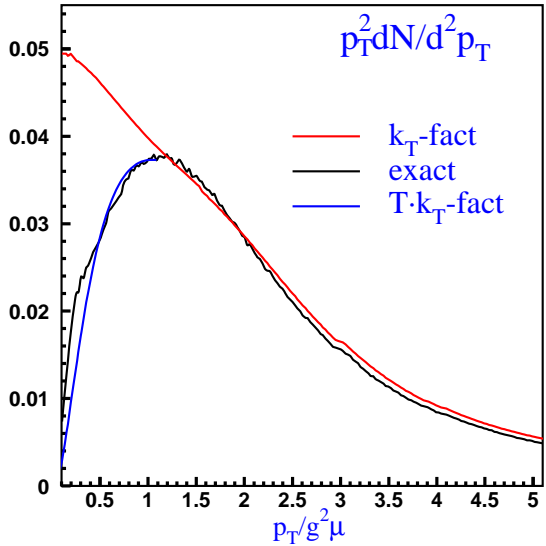}}
\end{minipage}
\begin{minipage}{7cm}{

\caption{The single inclusive cross section $p^2_T d N/d^2 p_T$ versus $p_T$. The low curve shows the exact solution given in Ref. \cite{BLMT} . The upper curve presents the result for the inclusive cross sections from $k_T$-factorization. Both these curves are taken from Ref. \cite{BLMT} and we are very thankful to Tuomas Lappi who shares with us the data for these curves. The blue curve (the shortest one)
shows the result for \eq{F9} assuming that $\bas = 0.1$ and $Q_s = 1.5 g^2\mu$. 
}
\label{lowp}
}
\end{minipage}
\end{figure}

In conclusions, we would like to summarize that we suggest \eq{F9} which violates the $k_T$ factorization for $k_T < Q_{s, 1}\, \mbox{and} \,Q_{s,2}$  but it is  in a perfect agreement with the numerical solution for the inclusive production in Colour Glass Condensate \cite{BLMT}( see \fig{lowp}). We would like to emphasize that our result is based on  general grounds in QCD and
reflects the fact that emission of soft gluons with transverse momenta $Q_s\,>\,p_{i,T}\, > \,p_T$ has been taken into account in functions $\phi_G$ (see  \eq{F9}) and it should not be included again in $k_T$-factorization formula.

\section*{Acknowledgements} 
We are very thankful to Yura Kovchegov whose paper \cite{KOV} stimulated hot discussions that led to this notes and who draw our attention to Ref.\cite{BLMT} which we, unfortunately, overlooked.  We would like also to thank Amir Rezaeian, who made \fig{num} for us, and Tuomas Lappi who shares with us the data for \fig{lowp}.  This work was supported in part by the  Fondecyt (Chile) grant  \# 1100648.

\end{document}